\documentclass[11pt]{article}
\usepackage[english]{babel}
\input{epsf}
\usepackage{mathtext}
\usepackage{graphicx}
\usepackage{graphics,color}
\usepackage[pdftex]{hyperref}
\newcommand{\hhh}{{\cal H}}
\newcommand{\CA}{{\cal A}}
\newcommand{\CB}{{\cal B}}
\newcommand{\CR}{{\cal R}}

\newcommand{\CN}{{\cal N}}
\newcommand{\CD}{{\cal D}}
\newcommand{\be}{\begin{equation}}
\newcommand{\ene}{\end{equation}}
\newcommand{\ba}{\begin{array}}
\newcommand{\ea}{\end{array}}

\newcommand{\bsigma}{\mbox{\boldmath$\sigma$}}

\newcommand{\btau}{\mbox{\boldmath$\tau$}}

\begin{document}

\title{Spin-triplet $f$-wave symmetry in superconducting monolayer $MoS_2$}
\author{H. Goudarzi$^1$\footnote{Corresponding author: h.goudarzi@urmia.ac.ir ; goudarzia@phys.msu.ru}, M. Khezerlou$^1$, H. Sedghi$^{1,2}$, A. Ghorbani$^2$\\
\footnotesize\textit{$^1$Department of Physics, Faculty of Science, Urmia University, Urmia, P.O.Box: 165, Iran}\\
\footnotesize\textit{$^2$Department of Basic Science, Urmia University Pardis, Urmia, Iran}}
\date{}
\maketitle

\begin{abstract}
The proximity-induced spin-triplet $f$-wave symmetry pairing in a monolayer molybdenum disulfide-superconductor hybrid features an interesting electron-hole excitations and also effective superconducting subgap, giving rise to a distinct Andreev resonance state. Owing to the complicated Fermi surface and momentum dependency of $f$-wave pair potential, monolayer $MoS_2$ with strong spin-orbit coupling can be considered an intriguing structure to reveal the superconducting state. Actually, this can be possible by calculating the peculiar spin-valley polarized transport of quasiparticles in a related normal metal/superconductor junction. We theoretically study the formation of effective gap at the interface and resulting normalized conductance on top of a $MoS_2$ under induction of $f$-wave order parameter, using Bloder-Tinkham-Klapwijk formalism. The superconducting excitations shows that the gap is renormalized by a \textit{bias limitation} coefficient including dynamical band parameters of monolayer $MoS_2$, especially, ones related to the Schrodinger-type momentum of Hamiltonian. The effective gap is more sensitive to the n-doping regime of superconductor region. The signature of spin-orbit coupling, topological and asymmetry mass-related terms in the resulting subgap conductance and, in particular,  maximum conductance is presented. In the absence of topological term, the effective gap reaches to its maximum value. The unconventional nature of superconducting order leads to the appearance of zero-bias conductance. In addition, the maximum value of conductance can be controlled by tuning the doping level of normal and superconductor regions.  
\end{abstract}

\section{Introduction}

Recently discovered monolayer molybdenum disulfide (ML-MDS) \cite{1,2} including two-dimensional (2D) Dirac-like charge carriers may present itself as a capable structure to demonstrate distinct transport properties resulted from Andreev-Klein process. Among peculiar dynamical properties of monolayer $MoS_2$, one can point out to: i) existing two inequivalent nondegenerate $K$ and $K'$ valleys originated from strong spin-orbit coupling (SOC)($\approx 0.08 \ eV$) in valence band, ii) appearance of a significant direct band gap in low-energy band structure in the visible frequency range $\approx 1.9 \ eV$, iii) inversion symmetry breaking caused by valence band valley-contrasting spin-splitting $\approx 0.1-0.5 \ eV$ \cite{8,10,12}.
In addition above dynamical features, layered structure of ML-MDS, chemical stability and relatively high mobility (room temperature mobility over $200\; cm^2/Vs$) can make it potentially a useful material for electronic applications \cite{9,11}. On the other hand, it has been experimentally shown that the superconductivity may be induced to the $MoS_2$ layer in the presence of a superconducting electrode near it via the proximity effect \cite{13,14,15,16}. Consequently, regarding the fact that charge carriers behave as Dirac-like nonzero effective mass particles (with Fermi velocity of $v_F\approx 0.53\times 10^6\; ms^{-1}$), predicted by Beenakker \cite{17,18} specular Andreev reflection (AR) can be realized at the ML-MDS normal metal/superconductor (NS) interface. Consequence of the existence of relativistic electron-like and hole-like quasiparticles can be understood by studying tunneling conductance, resulted from Andreev-Klein process in a NS junction. In this regard, there are several investigations by many authors \cite{19,20,24,36,44,45,21,22,23}.

However, the hexagonal symmetry of the ML-MDS lattice permits for unconventional superconducting order parameters via the proximity induction. In particular, spontaneous time-reversal symmetry breaking in spin-triplet pairing, the several anomalous transport phenomena such as the anomalous Hall effect, polar Kerr effect for microwave radiation, anomalous Hall thermal conductivity, and anomalous photo-and acousto-galvanic effects may be exhibited in the absence of external magnetic fields \cite{25,26,27}. Recently, some authors have also explored the chiral $p$-wave pairing in the various systems leading to anomalous transport phenomena and surface states \cite{27,28}. An intense research activity has been allocated to the possibility to induce a spin-triplet state using a conventional $s$-wave superconductivity in the last few years. This, actually, can play an essential role in 2D materials with strong spin-orbit coupling \cite{M1,M2,M3,M4}.  Moreover, it has been shown that the Majorana bands can be appeared in the vortex state of a chiral $p$-wave superconductor \cite{29}. Triplet pairing may also give rise to suppress the superconducting surface states in topological insulators \cite{30,31,32,33,34}. Regarding our previous works \cite{21,35}, in this paper, we analytically investigate the signature of spin-triplet $f$-wave symmetry in low-energy excitations of Dirac-like charge carriers in ML-MDS. $Sr_2RuO_4$ and $UPt_3$ are among the materials that the superconductivity in them seems to be carried by spin-triplet pair potential \cite{37,38,39}, and the particular possible $f$-wave symmetry is heavy fermions complex $UPt_3$ \cite{40}. Another possibility is quasi-one dimensional organic system \cite{41}. The Fermi surface of this order parameter is more complicated, and the pairing mechanism could not yet been determined exactly. 

Electron-hole exchange at the interface of a normal/superconductor in quasiparticle excitations below superconducting gap can be affected by the off-diagonal elements of spin-triplet Bogoliubov-de Gennes gap matrix. The spin-triplet superconducting quasiparticle excitations in $MoS_2$ are found to play a crucial role in AR process, since the superconducting gap can be renormalized by electron(hole) wave vector and as well as chemical potential \cite{21}. Furthermore, in ML-MDS, the AR process is believed to be spin-valley polarized due to the valley-contrasting spin splitting in the valence band, and indeed, depends on the magnitude of the chemical potential (doping of N and S regions). More importantly, the electron-hole difference mass $\alpha$ parameter and also topological $\beta$ parameter \cite{43} relating to the Schrodinger-type kinetic energy are taken into the Dirac-Bogoliubov-de Gennes (DBdG) ML-MDS Hamiltonian. The contribution of these terms in superconducting electron-hole wave vector $k_{s}^{e(h)}$ gives rise to change significantly the AR and resulting subgap tunneling conductance results. This paper is organized as follows. Section 2 is devoted to the analytical solutions of the ML-MDS DBdG equation inducing $f$-wave order parameter to obtain the exact expressions of dispersion energy and corresponding Dirac spinors. We represent the explicit expressions of normal and Andreev reflection amplitudes in a corresponding NS junction. The numerical results of effective gap and resulting subgap conductance are presented in Sec. 3 with a discussion of main characteristics of system. Finally, a brief conclusion is given in Sec. 4.

\section{Theoretical formalism}
\subsection{Spin-triplet ML-MDS superconductor}

The low-energy fermionic excitations at the two distinct Dirac points in Brillouin zone of monolayer $MoS_2$ can be obtained by consideration of relativistic massive quasiparticles including spin-orbit interaction and Schrodinger-type kinetic energy term. Near these points, the electronic dispersion is described by the 2D Dirac equation and the corresponding Hamiltonian, in addition to linear dependence on momentum contains the quadratic terms originated from the physical sense related to the difference mass between electron and hole and topological characteristic parameters. The strong SOC leads to distinct spin-splitting energy in the valence band for two different $K$ and $K'$ valleys (see, Fig. 1(b)). Hence, the effective Hamiltonian of ML-MDS around the Dirac points reads:
\begin{equation}
\hat{\hhh}=\hbar v_{F}\hat{\mathbf{k}}\cdot\hat{\bsigma}_{\tau}+\Delta\hat{\sigma}_{z}+\lambda s\tau(\frac{\hat{I}_2-\hat{\sigma}_{z}}{2})+\frac{\hbar^{2}|k|^2}{4m_0}\left(\alpha\hat{I}_2+\beta\hat{\sigma}_{z}\right),
\end{equation}
where $\bsigma_{\tau}=(\tau\sigma_{x},\sigma_{y})$ are the Pauli matrices. The spin up and down is labeled by $s=\pm 1$, and valley index $\tau =\pm 1$ denotes the $K$ and $K'$ valleys. $\Delta$ and $m_0$ denote the direct band gap of ML-MDS and bare electron mass, respectively. The topological and mass difference band parameters are evaluated by $\beta =2.21$ and $\alpha =0.43$, respectively. $\lambda\approx 0.08\; eV$ and $v_F$ represent the SOC and Fermi velocity of charge carriers, respectively. By diagonalizing the Eq. (1), the excitation energy relative to the Fermi energy can be obtained as:
\begin{equation}
\epsilon=\left|\lambda s \tau+\frac{\hbar^{2}\left|k\right|^2}{2m_0}(\frac{\alpha}{2})+\eta\sqrt{\left(\Delta-\lambda s \tau+\frac{\hbar^{2}\left|k\right|^2}{2m_0}(\frac{\beta}{2})\right)^2+v^{2}_{F}\hbar^{2}\left|k\right|^{2}}\right|,
\end{equation}
where the parameter $\eta=\pm 1$ refers to the excitations in the conduction and valence bands. The superconducting electron-hole quasiparticles caused by the proximity induction of a superconductor electrode on top of $MoS_2$ monolayer can be realized by the generalization of Bogoliubov-de Gennes Hamiltonian. The resulting Dirac-Bogoliubov-de Gennes Hamiltonian for the triplet components is given by:
\begin{equation}
\hat{H}_{DBdG}=\left(\begin{array}{cc}
\hat{\hhh}-E_{F}+U(x)&\hat{\Delta}_{s}(\textbf{k})\\
\hat{\Delta}_{s}^{\ast}(\textbf{k})&-\hat{\hhh}+E_{F}-U(x)
\end{array}\right),
\end{equation}
where the electrostatic potential $U(x)$ gives the relative shift of Fermi energy, which is determined by the magnitude of the chemical potential; $\mu_s=E_{F}-U(x)$. The symmetry of a Cooper pair order is constructed by application of the time-reversal, and charge conjugation operators. Generally, the possible pairing states can be traditionally characterized as spin-singlet with orbital even parity or spin-triplet with odd parity order parameters. The superconducting order parameter in the spin-triplet symmetry is  parameterized by the $\textbf{d}(\textbf{k})$ vector (its direction is perpendicular to the total spin of a Cooper pair) as $\hat{\Delta}_s(\mathbf{k})=\left[\mathbf{d}(\mathbf{k})\cdot \btau \right]i\tau_y$, where $\mathbf{d}(\mathbf{k})=\Delta_f(\textbf{k}) \hat{z}$, $\btau$, and $\tau_y$ are an odd-parity function of $\mathbf{k}$ and Pauli matrices, which describe the real electron spin, respectively.
The order parameter for triplet unconventional superconductors with $f$-wave symmetry has a more complex dependence on $\mathbf{k}$. Its spatial orbital form can be a contorted surface, as shown in Fig. 1(c). We notice that the ML-MDS lattice is hexagonal, and the possible types of superconductivity that is induced by the electron-electron and electron-phonon interactions has been discussed in Ref. \cite{46}. Here, the momentum dependence of $f$-wave pair potential may be written as \cite{48,49}:
$$\Delta_f(\textbf{k})=\Delta_{0f}\cos{\theta_s}\left(\cos^2{\theta_s}-3\sin^2{\theta_s}\right),$$
where $\theta_s$ and $\Delta_{0f}$ are the angle of incidence of electron to the interface and net value of pair potential, respectively. Owing to the high orbital quantum number, the spin-triplet $f$-wave superconducting gap may be, of course, of considerable influence in 2D materials with strong spin-orbit coupling. However, the wavevector dependence of pair potential on the surface shows the maximum gap for $k_x/k_y=\pm 1$. The nodal points of gap appear in only $k_x=0$ or $k_x/k_y=\pm \sqrt{3}$. Even if we set $k_y=0$, the net gap $\Delta_{0f}$ does not become zero for $k_x\neq 0$. The gap changes its sign when $k_x\rightarrow -k_x$, whereas we have no change in sign of superconducting gap by inversion of $k_y$. These are shown in Fig. 2. It is important that, for superconducting hole excitations this gap may change the sign, since we have $\theta^h_s=\pi-\theta^e_s$ and, thus, $k_x\rightarrow -k_x$. This feature is an interesting signature of unconventional anisotropic superconductivity, giving rise to the zero-bias conductance \cite{ZES}. These features confirm a more complicated pair potential of $f$-wave symmetry.

The DBdG Hamiltonian may be diagonalized to yield the dispersion relation of ML-MDS superconducting electron-hole excitations:
\begin{equation}
\epsilon_{S}=\xi\sqrt{\left(\frac{\CA+\CB}{2}+\nu\sqrt{\left(\frac{\CA-\CB}{2}\right)^2+\hbar^{2}v^{2}_{F}\left|k_{s}\right|^{2}+(1-\eta^2)\left|\Delta_{f}(\textbf{k})\right|^2}\right)^2+\eta^2\left|\Delta_{f}(\textbf{k})\right|^2},
\end{equation} 
where the parameter $\xi=\pm 1$ denotes the electron-like and hole-like excitations, while $\nu=\pm 1$ distinguishes between the conduction and valence bands, which is demonstrated in Fig. 1(b). Note that, the magnitude of the superconducting order parameter is now renormalized by the coefficient
$$\eta=\sqrt{1-\left(\frac{\CA-\CB}{\CA+\CB}\right)^2} ,$$
where
$$\CA=\Delta+\frac{\hbar^{2}\left|k_s\right|^2}{4m_0}(\alpha+\beta)-\mu_s , \  \CB=-\Delta+2\lambda s \tau+\frac{\hbar^{2}\left|k_s\right|^2}{4m_0}(\alpha-\beta)-\mu_s$$
which contains the dynamical characteristic of ML-MDS, and as well the chemical potential. It is obvious that the opened in Fermi level effective superconducting gap can be suppressed by the parameter $\eta$, which is tuned by band parameters of $MoS_2$. Thus, the tunneling subgap conductance may be influenced by the tuning of above band feature of ML-MDS. The energy-momentum relation can be solved to obtain the exact form of the superconducting electron(hole) wavevector:
$$
k^{e(h)}_s=\frac{1}{\hbar v_F}\left(k_0+ik_1\right) ; \ \ \ k_0=\sqrt{\CA\CB},
$$
where $k_1$ may be responsible to the exponentially decaying.

\subsection{NS junction}

In this section, we proceed to investigate the effect of induced spin-triplet $f$-wave symmetry superconducting order on the transport properties of a NS junction, where N region is in $x<0$, and S region is in $x\geq 0$, as sketched in Fig. 1(a). Taking the superconducting gap to be zero in N region, the eigenstates of Eq. (3) can be obtained as Dirac spinors for electron and hole excitations. Denoting the amplitude of normal and Andreev reflections, respectively, by $\CR(\theta_N^e,\epsilon)$ and $\CR_A(\theta_N^e,\epsilon)$, the total wave function for a right-moving electron with angle of incidence $\theta^{e}_{N}$, a left-moving electron by the substitution $\theta^{e}_{N}\rightarrow \pi-\theta^{e}_{N}$, and a left-moving hole by angle of reflection $\theta^{h}_{N}$ can be described by:
$$
\psi_{N}=e^{ik_{y}y}\left[\frac{1}{\sqrt{\CN_{e}}}\left(1\ \ \tau e^{i\tau\theta^{e}_{N}}\ \ 0\ \ 0\right)^{T}e^{i\tau k^e_{x}x}+\frac{\CR(\theta_N^e,\epsilon)}{\sqrt{\CN_{e}}}\left(1\ \ -\tau e^{-i\tau\theta^{e}_{N}}\ \ 0\ \ 0\right)^{T}e^{-i\tau k^e_{x}x}+\right.
$$
\begin{equation}
\left.+\frac{\CR_A(\theta_N^e,\epsilon)}{\sqrt{\CN_{h}}}\left(0\ \ 0\ \ 1\ \ \tau e^{-i\tau\theta^{h}_{N}}\right)^{T}e^{i\tau k^h_{x}x})\right],
\end{equation}
where the normalization factor $\CN_{e(h)}$ ensures that the quasiparticle current density of states is the same. The conserved ($k_y=k\sin{\theta^{e}_{N}}$) and nonconserved ($k^{e(h)}_x=k\cos{\theta^{e(h)}_{N}}$) components of the momentum can be acquired from Eq. (2). The DBdG Hamiltonian can be solved to obtain the electron-like and hole-like quasiparticles eigenstates for superconductor region and the wavefunction including a contribution of both eigenstates are analytically found as \cite{21}:
\begin{equation}
\psi_{S}=t\left(\begin{array}{cc}
\zeta\beta_{1}\\
\zeta\beta_{1}e^{i\tau\theta_s}\\
e^{-i\gamma_{e}}e^{-i\varphi}e^{i\tau\theta_s}\\
e^{-i\gamma_{e}}e^{-i\varphi}
\end{array}\right)e^{i\tau k^e_{sx}x}+t'\left(\begin{array}{cc}
\zeta\beta_{2}\\
-\zeta\beta_{2}e^{-i\tau\theta_s}\\
-e^{-i\gamma_{h}}e^{-i\varphi}e^{-i\tau\theta_s}\\
e^{-i\gamma_{h}}e^{-i\varphi}
\end{array}\right)e^{-i\tau k^h_{sx}x},
\end{equation}
where we define
$$
\beta_{1(2)}=-\frac{\epsilon_S}{\left|\Delta_{f}\right|}-(+)\sqrt{\frac{\epsilon_S}{\left|\Delta_{f}\right|^2}-\eta^2},\ \ \zeta=\frac{\CA+\CB}{2\sqrt{\CA\CB}},\ \ e^{i\gamma_{e(h)}}=\frac{\Delta_{f}(\theta^{e(h)}_s)}{\left|\Delta_{f}\right|}. 
$$
Here $t$ and $t'$ correspond to the transmission of electron and hole quasiparticle, respectively. By applying the boundary condition for wavefunctions at the interface ($x=0$) of junction, the following solution for Andreev and normal reflection coefficients is found:
$$
\CR_A(\theta_N^e,\epsilon)=\sqrt{\frac{\CN_{e}}{\CN_{h}}}\CD\left(e^{i\tau\theta_s}e^{-i\gamma_{e}}\mathcal{M}_{4}-e^{-i\tau\theta_s}e^{-i\gamma_{h}}\mathcal{M}_{3}\right),
$$
\begin{equation}
\CR(\theta_N^e,\epsilon)=\CD\zeta\left(\beta_{1}\mathcal{M}_{4}+\beta_{2}\mathcal{M}_{3}\right)-1,
\end{equation}
where we have introduced
$$
\CD=\frac{2\tau\cos{\theta^{e}_{N}}}{\mathcal{M}_1\mathcal{M}_{4}-\mathcal{M}_{2}\mathcal{M}_{3}},
$$
$$
\mathcal{M}_{1(2)}=\zeta\beta_{1(2)}\left((-)\tau e^{-i\tau\theta^{e}_{N}}+e^{(-)i\tau\theta_s}\right),
$$
$$
\mathcal{M}_{3(4)}=e^{-i\gamma_{e(h)}}\left(\tau e^{-i\tau\theta^{e}_{N}}e^{(-)i\tau\theta_s}-(+)1\right).
$$
Finally, the tunneling subgap conductance $G (eV)$ passing through the N/S junction can now be calculated in terms of the reflection amplitudes via the Blonder-Tinkham-Klapwijk (BTK) formalism \cite{47}:
\begin{equation}
G(eV)=\sum_{s,\tau=\pm 1} G^{s,\tau}_0\int^{\theta_{c}}_{0}\left(1-\left|\CR(\theta_N^e,\epsilon)\right|^{2}+\left|\CR_A(\theta_N^e,\epsilon)\right|^{2}\right)\cos{\theta_{N}^e}d\theta_{N}^e,
\end{equation}
where $G^{s,\tau}_{0}=e^{2}N_{s,\tau}(eV)/h$ is the ballistic conductance of spin and valley-dependent transverse modes $N_{s,\tau}=kw/\pi$ in a sheet of $MoS_{2}$ of width $w$, that $eV$ denotes the bias voltage. The upper limit of integration in above equation needs to obtain exactly based on the fact that the incidence angle of electron-hole at the interface may be less than a value of $\pi/2$.

\section{RESULTS AND DISCUSSION}
\subsection{Effective subgap}

We now proceed to investigate the transport properties of a NS junction, when the $f$-wave symmetry pairing is deposited on top of a ML-MDS. First, we consider the behavior of superconducting effective gap with respect to the incident electron from N region with angle $\theta^e_N$ to the interface. Indeed, this may determine the exhibition of Andreev process and resulting subgap tunneling conductance. In one-dimensional limit, transport of tunneling electrons is in the $x$-direction with wavevector $k_x$ and incident angle $0\leq\theta_N^e\leq\pi/2$, and conservation of momentum in $y$-direction enables us to find superconductor incident angle $\theta_s$ in terms of $\theta_N^e$, which is $\theta_s=\sin^{-1}\left(\frac{k}{k_s}\sin\theta_N^e\right)$. On the other hand, the coefficient $\eta$ appeared in superconducting excitations of Eq. (4) couples with pair potential and, consequently, gives rise to reduce the effective subgap \cite{21}. In our system, the mean-field approximation necessitates the chemical potential of S region to be much larger than the pair potential $\mu_s\gg\Delta_{0f}$, for low magnitude of $\mu_s$, the effective subgap notably decreases with increase of angle of incident electron to the NS interface. The reason can be described by the strong spatial dependency of $f$-wave symmetry pairing. We demonstrate this effect in Fig. 3(a), which is in contrast to the spin-triplet $p$-wave symmetry. We plot the normalized effective subgap $\Delta_f(\textbf{k})/\Delta_{0f}$ taking into account the \textit{bias-limitation} coefficient $\eta$, including the dynamical features of ML-MDS, e.g. SOC parameter $\lambda$ and topological term $\beta$. If we set $\eta=1$, then the maximum of $\Delta_f(\textbf{k})/\Delta_{0f}$ becomes unit for normal electron incident angle, $\theta^e_N=0$. This is an important point, that it needs to more accurately consider calculating subgap conductance. 

Fig. 3(a) shows that, for higher magnitude of $\mu_s$, the effective gap presents no dependency on electron incident angle, owing to the result $\theta_s\cong 0$, and a slowly decreasing behavior. The above results have been achieved for the actual case of ML-MDS. We now examine the $f$-wave effective subgap in the absence of band parameters of monolayer $MoS_2$. We present in Fig. 3(b) the effect of topological term ($\beta$) contribution, which is dominant in higher superconductor chemical potentials and responsible for the suppression of the effective gap. In the absence of $\beta$, the effective gap reaches its maximum value, which is $\Delta_f(\textbf{k})/\Delta_{0f}=1$. If we ignore the SOC ($\lambda$) contribution, the effective gap does not close even if for parallel incident to the interface in low doping S region. This effect is demonstrated in Fig. 3(c), which argues that the superconducting ML-MDS with strong SOC can be affected by the spin-triplet pair potential owing to the momentum dependence of order parameter interacting with spatial orbital momentum of system. Finally, we show in Fig. 3(d) the asymmetry mass term $\alpha$, originated from the electron-hole difference mass, dependence of effective gap. As reported in previous works \cite{21,43}, we see that the mass-related band parameter of ML-MDS has almost no effect on the effective subgap. Since, the band parameters $\alpha$ and $\beta$ are related to the Schrodinger-type wave vector of Dirac-like charge carriers in ML-MDS, the resulting tunneling subgap conductance can be controlled by these parameters, which will be investigated and discussed in the next section.   

\subsection{Conductance spectroscopy}

Our proposed junction, sketched in Fig. 1(a), is an ideal setup to measure the conductance spectroscopy like the ones recently performed in Refs. \cite{45,21,22}. By using Eq. (8), we calculate numerically the normalized conductance $G/G_0$, as it is commonly done in experiments. Regarding the band structure of ML-MDS, and in order to have propagating states in N region we choose the p-doped case, that the chemical potential is limited to be in range $-1.1\leq\mu_N\leq -0.94 \ eV$, due to valence band spin-splitting. Furthermore, the effective gap curves (see, Fig. 3) predicates the normalized bias-energy of junction $\epsilon(eV)/\eta|\Delta_f|$ to reduce, comparing to its common value $\epsilon(eV)/|\Delta|=1$. Figure 4 demonstrates the subgap conductance of the system resulting from the Andreev process. For the chemical potential of S region $\mu_s=5 \ eV$, the influence of dynamical characteristics of ML-MDS in conductance is presented. For the case of p-doped S region, we find that the effective gap reaches a negative value. Therefore, the p-doped S region regime is actually forbidden in this system. This is in sharp contrast with $p$-wave symmetry case, where we are allowed to be in the p-doped S region \cite{21}. 

Importantly, the zero-bias conductance (ZBC) is expected to appear owing to midgap resonant states, which is discussed in the following. A sharp conductance dip is obtained at effective gap $\epsilon(eV)/\eta|\Delta_f|=0.23$, and the resulting conductance becomes zero. The absence of spin-orbit coupling results in an increase of ZBC and, on the other hand, change the place of zero-conductance dip, occurring in lower bias. This can be realized by the fact of the nature  of spin-triplet superconductivity interplaying with SOC. Fig. 4 shows that the topological band parameter $\beta$ significantly affects the ZBC, whereas the mass-related term $\alpha$ has no substantial effect on the spin-valley polarized charge transport. Formally, consequence of ZBC can be described by the fact that it originates from the Andreev resonance energy, forming at the NS interface. This energy can be achieved when we choose the normal reflection to be zero. To more clarify this situation, we present, in Fig. 5, the Andreev resonance state (ARS) versus incident angle of electron to the interface for various n-doping of S region. For normal incidence, we see that the slope of ARS curve varies slightly with increase of the chemical potential. Therefore, ZBC may be enhanced for high n-doped S region. The slope of angle-resolved ARS has a zero value in the parallel incidence ($\theta_N^e=\pm\pi/2$).
Strikingly, the special $f$-wave symmetry results in a dominant energy distribution for low n-doped S region, $\mu_s=1.5 \ eV$, which has not been observed in $p$-wave symmetry case \cite{21}. 

To explore, in detail, the effect of chemical potential of N and S regions, we proceed to plot the maximum of normalized conductance as a function of $\mu_s$ for various p-doped chemical potential of N region. Owing to the spin-splitting gap in the valence band, for lower p-doping, we find the maximum of conductance to show increasing, notably. While, the $G_{Max}/G_0$ decreases in higher n-doped S region, as shown in Fig. 6(a). Hence, it can be possible to control the conductance maximum by tuning the level of p- or n-doping of N or S region, respectively. Finally, if we assume that the essential band parameters ($\beta$ and $\lambda$) of ML-MDS can be tunable (for example, in monolayer $MoS_2$ under strain, or in other similar 2D Dirac materials), then we are able to control the tunneling subgap conductance. In this regard, the SOC and topological parameters dependence of maximum conductance are demonstrated, respectively in Figs. 6(b) and (c), where the $G_{Max}/G_0$ decreases with the increase of SOC strength in the case of any n-doping S region. The influence of topological band parameter $\beta$ in maximum conductance is presented in Fig. 6(c).

\section{CONCLUSION}

In this paper, we have analyzed the effect of proximity-induced $f$-wave symmetry pairing in a normal-superconductor hybrid junction on top of a ML-MDS. The spatial momentum dependence of this pair potential is complicated. The hexagonal symmetry of ML-MDS lattice permits for unconventional order parameters. We have found that the spin-triplet monolayer $MoS_2$ superconducting excitations is more complicated, so that, the effective superconducting subgap for incident electrons from N region to the interface is strongly affected by the dynamical band parameters of ML-MDS. In the n-doped S region regime, the effective gap significantly depends on the chemical potential, and as well incident angle of electrons. We have investigate the influence of strong spin-orbit coupling $\lambda$, topological term $\beta$ contributions, and also level of doping of N and S regions in the tunneling subgap conductance. The maximum value of normalized conductance has been found to control by tuning the above physical parameters of system. In particular, we have obtained the maximum of conductance to appear in the case of low p-doped N region. The symmetric Andreev resonance state with respect to the electron incident angle to a NS interface for $-\pi/2\leq\theta_N^e\leq\pi/2$ has been achieved.

\newpage

\textbf{Figure captions}\\
\textbf{Figure 1(a), (b), (c)} (color online) (a) Sketch of a normal-superconductor junction on the monolayer $MoS_2$ with $f$-wave pairing symmetry. Superconductivity is induced in the system by proximity effect. The chemical potential in normal and superconductor regions can be controlled. (b) The energy dispersion of ML-MDS that spin-up and spin-down subbands of the valence band are separated by the strong spin-orbit coupling (left panel). The large band gap separates the valence and conduction bands and anisotropic superconducting gap opens in the Fermi level (right panel). (c) Sketch of $f$-wave order parameter profile in the spherical coordinate.\\
\textbf{Figure 2} (color online) Contour plot of the $f$-wave order parameter, clearly showing the nodal line of the superconducting gap. It is seen that the maximum gap occurs in $k_x/k_y=\pm 1$.\\
\textbf{Figure 3(a), (b), (c), (d)}(color online) Effective band gap resulting from the dynamical characteristic of ML-MDS with $\mu_N=-0.96\ eV $ as a function of n-doped superconductor region and angle of incident electron from N region. (a) Change of gap anisotropy in the $p$-wave (brown surface) and $f$-wave symmetries (purple surface) that we set $\alpha=0.43$, $\beta=2.21$ and $\lambda=0.08\ eV$. (b) Dependence of the effective band gap for two different values of $\beta=0$ (brown surface) and $\beta=2.21$ (purple surface) when $\alpha=0.43$ and $\lambda=0.08\ eV$. (c) Dependence of the effective band gap for two different values of $\lambda=0$ (brown surface) and $\lambda=0.08$ (purple surface) when $\alpha=0.43$ and $\beta=2.21$. (d) Dependence of the effective band gap for two different values of $\alpha=0$ (brown surface) and $\alpha=0.43$ (purple surface) when $\lambda=0.08$ and $\beta=2.21$.\\
\textbf{Figure 4}(color online) Dependence of the tunneling subgap conductance on the bias voltage for several values of $\alpha,\beta,\lambda$, when $\mu_s=4\ eV$ and $\mu_N=-1.1\ eV$.\\
\textbf{Figure 5}(color online) The Andreev resonance energy as a function of the electron incident angle for several values of superconductor chemical potential, when $\alpha=0.43$, $\beta=2.21$, $\lambda=0.08$. and $\mu_N=-0.96\ eV $.\\
\textbf{Figure 6(a), (b), (c)}(color online) The maximum of normalized conductance as a function of (a) chemical potential of S region (b) spin-orbit coupling term and (c) topological band parameter. we have set $\mu_N=-1.1\ eV$ in Figs (b) and (c).\\ 

\newpage

\begin{figure}[ht]
\centering
\includegraphics[scale=0.8]{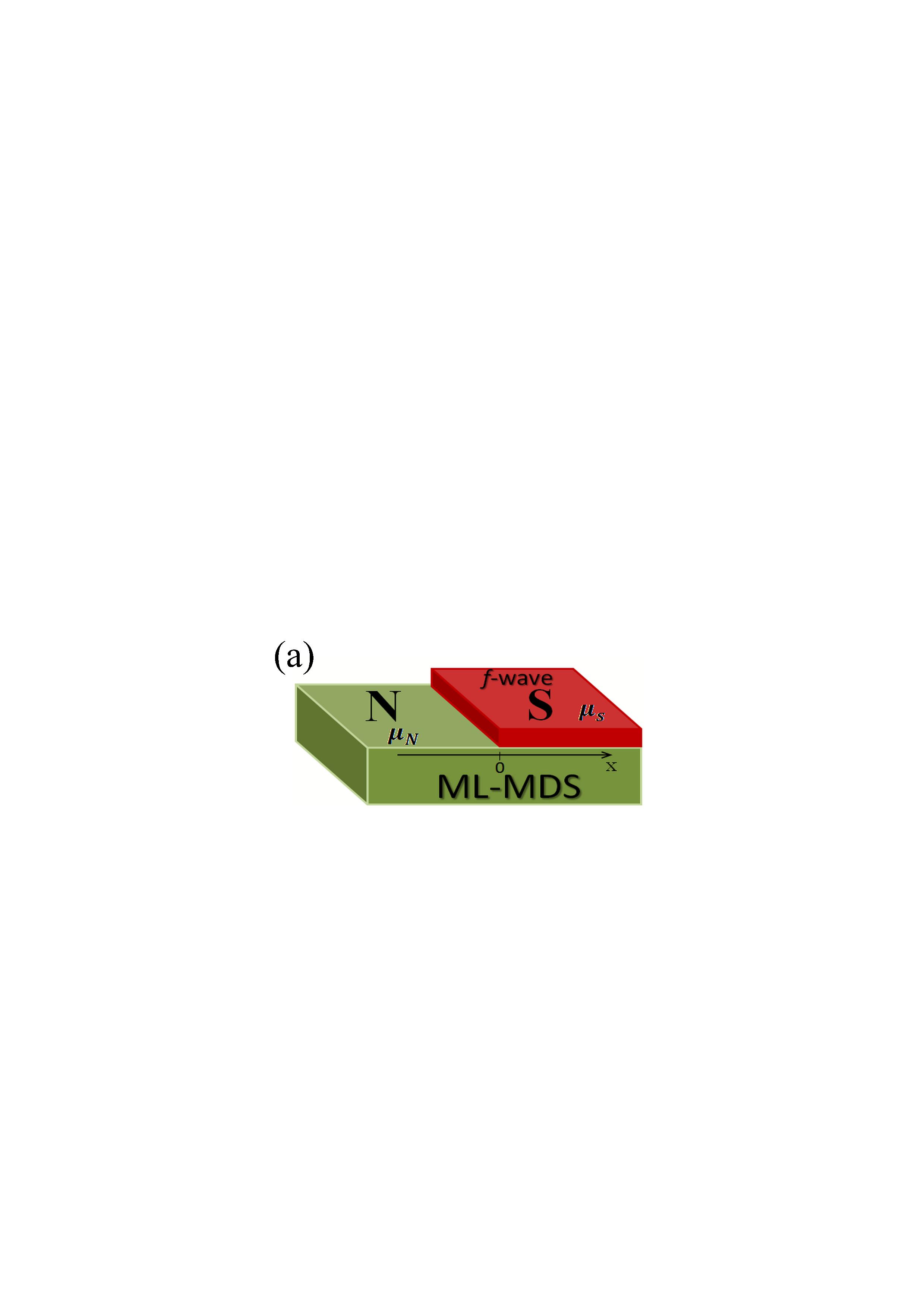}
\end{figure}

\begin{figure}[ht]
\centering
\includegraphics[scale=0.8]{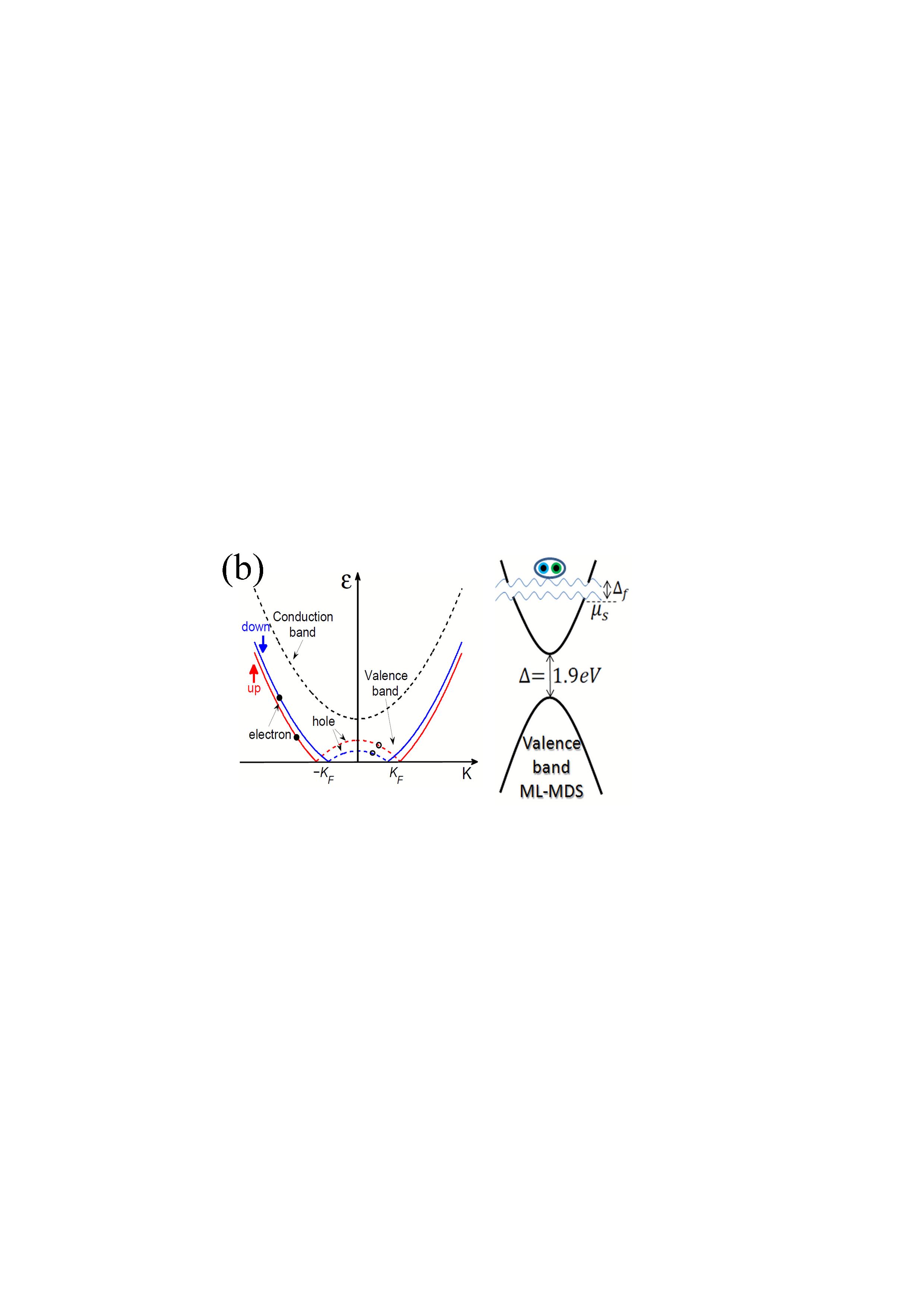}
\end{figure}

\begin{figure}[ht]
\centering
\includegraphics[scale=0.6]{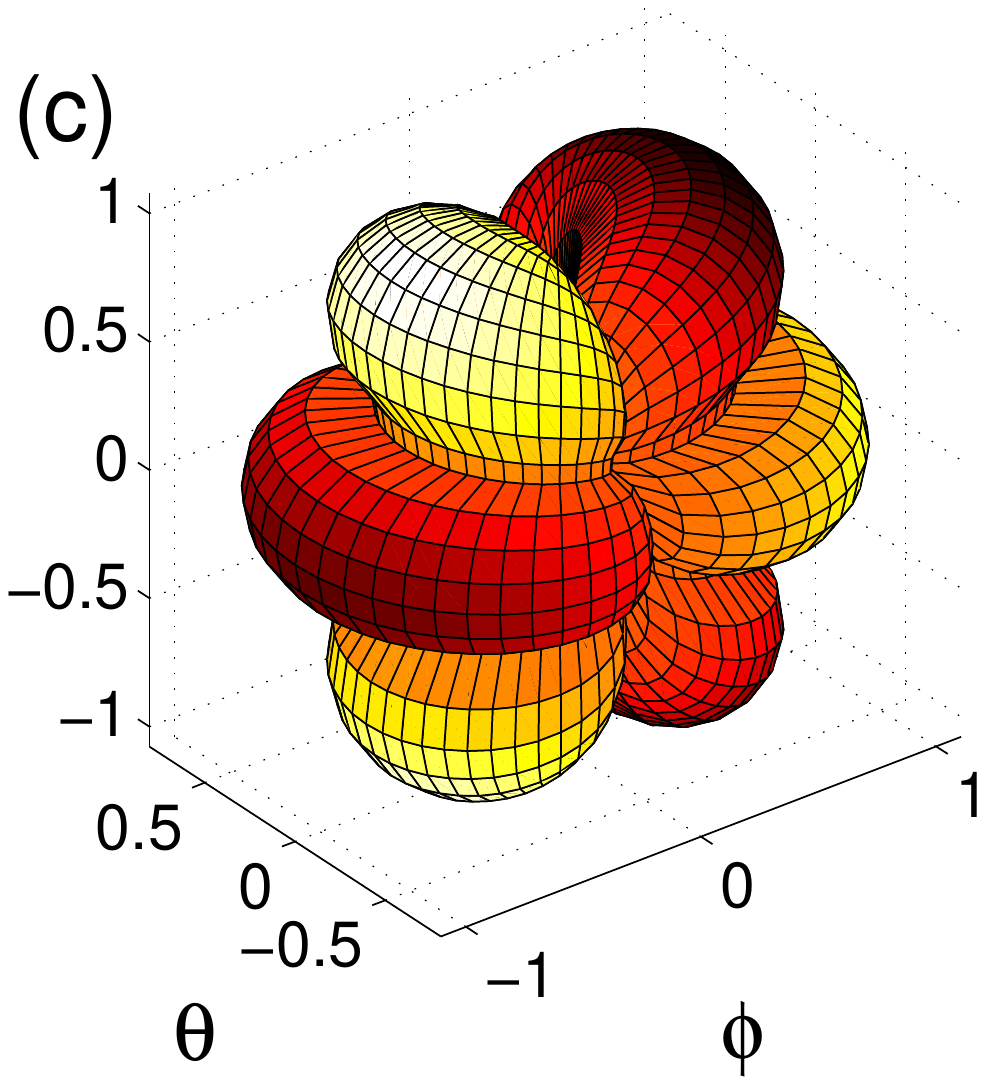}
\caption{(a),(b),(c)}
\label{fig1}
\end{figure}

\begin{figure}[ht]
\centering
\includegraphics[scale=0.5]{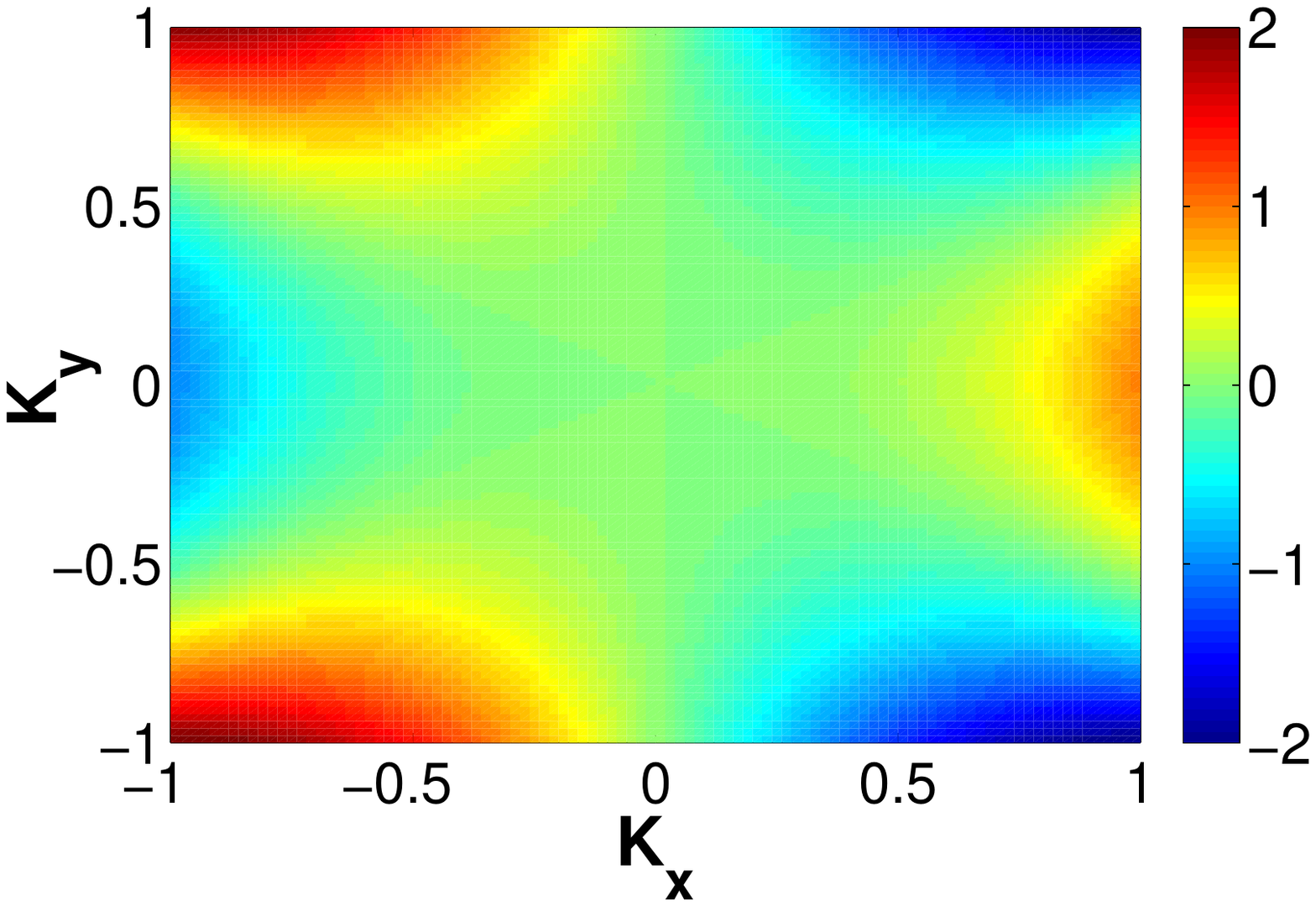}
\caption{}
\label{fig2}
\end{figure}

\begin{figure}[ht]
\centering
\includegraphics[scale=0.4]{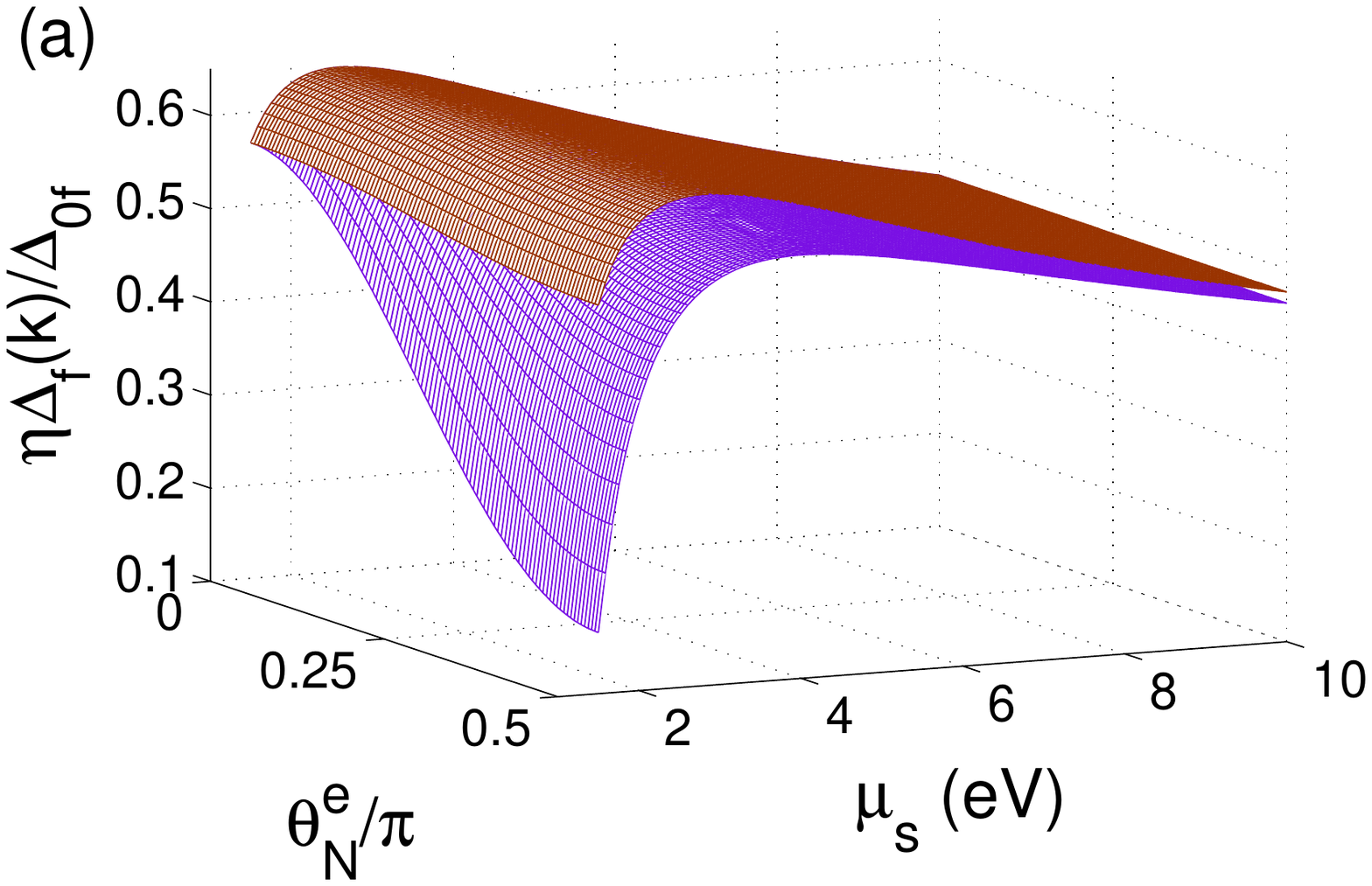}
\end{figure}

\begin{figure}[ht]
\centering
\includegraphics[scale=0.4]{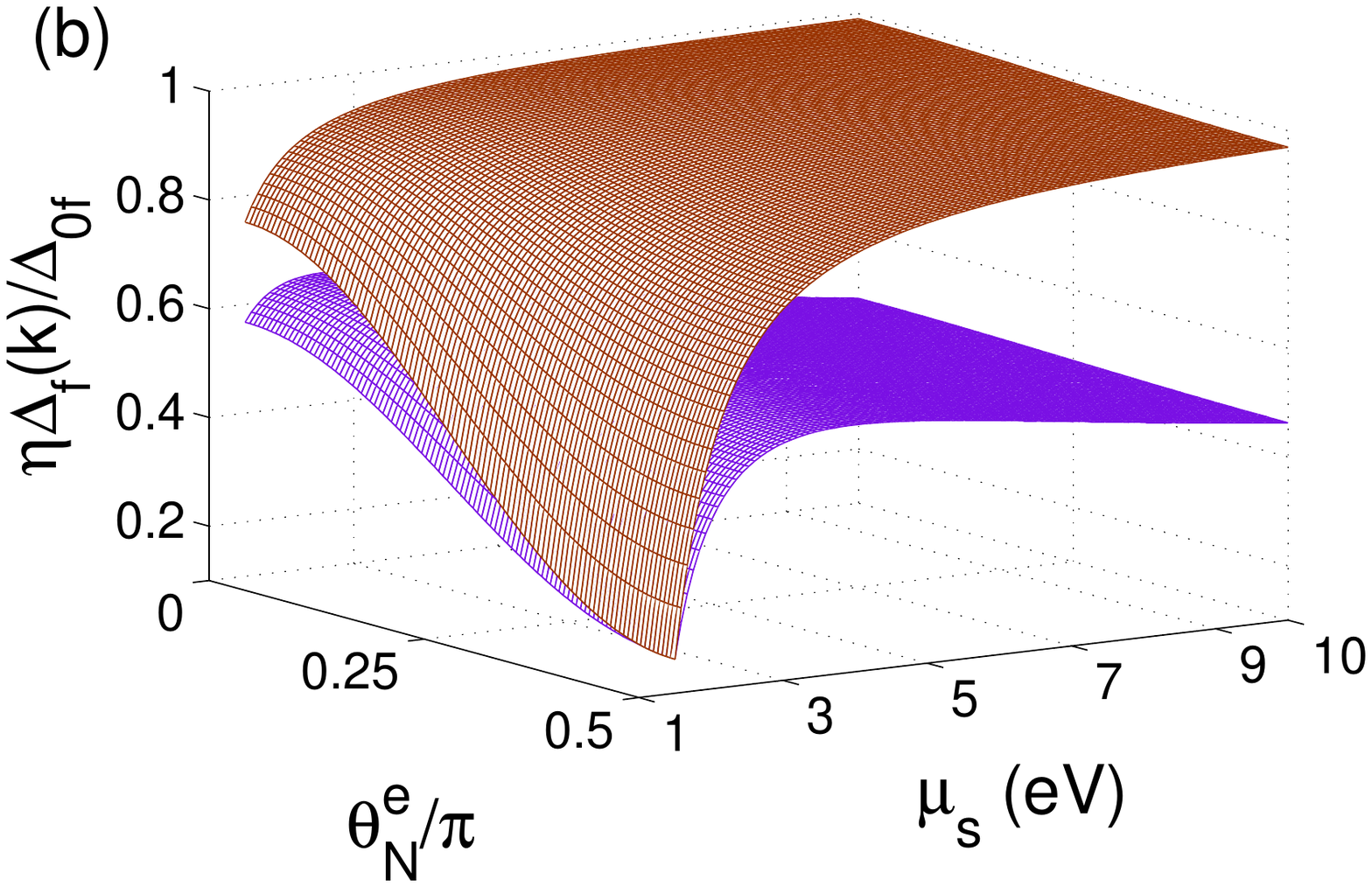}
\end{figure}

\begin{figure}[ht]
\centering
\includegraphics[scale=0.4]{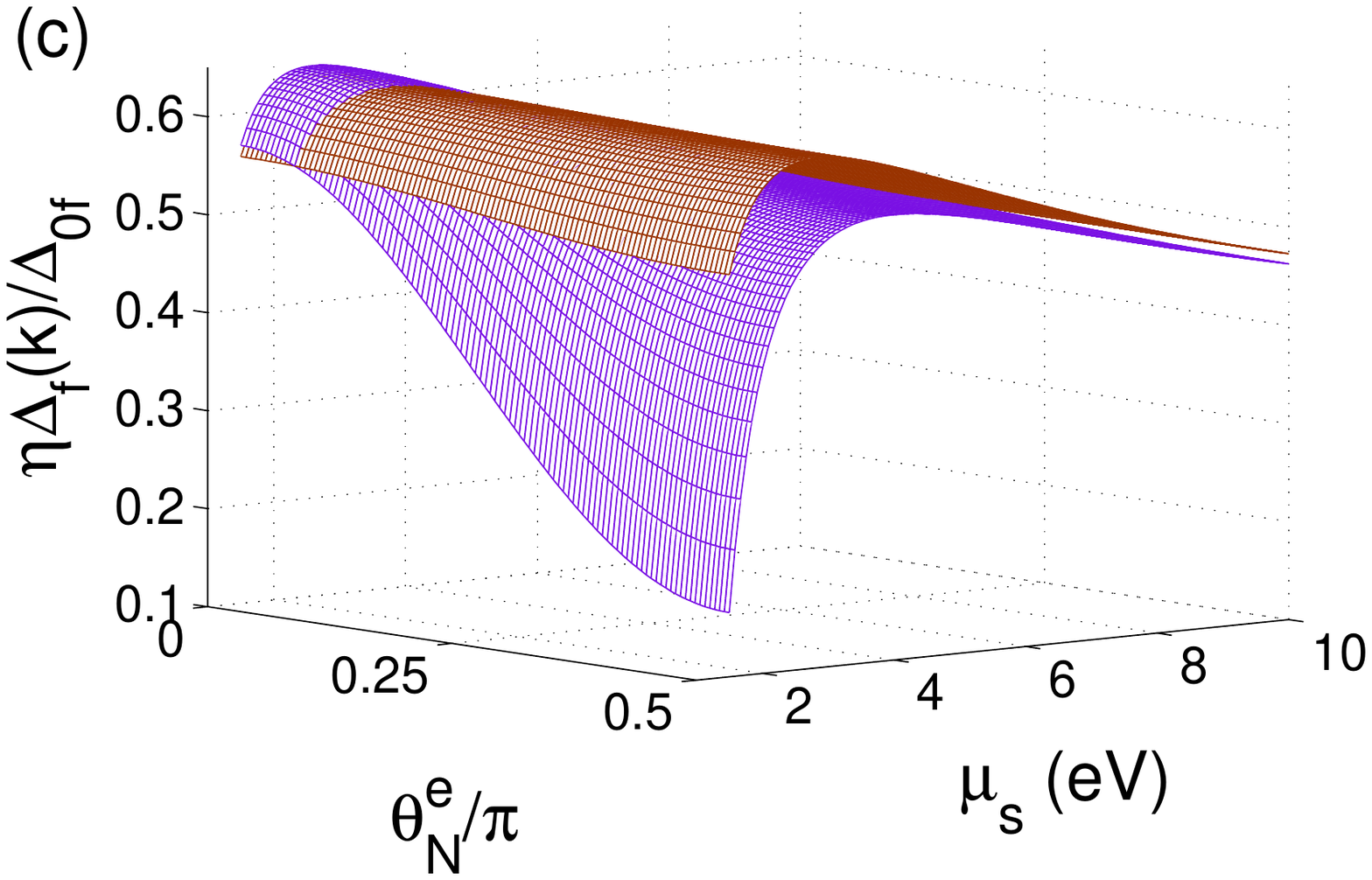}
\end{figure}

\begin{figure}[ht]
\centering
\includegraphics[scale=0.4]{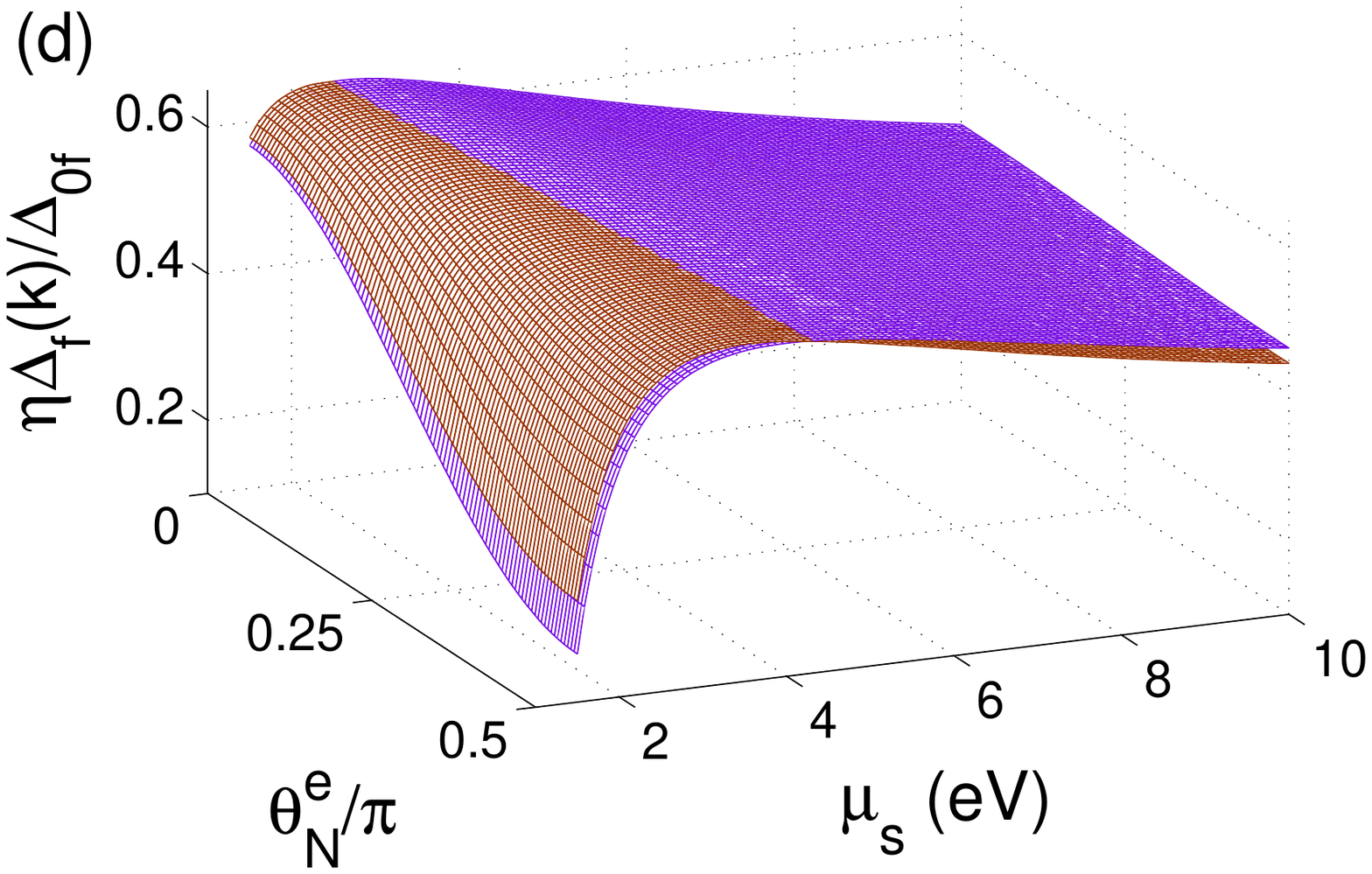}
\caption{(a),(b),(c),(d)}
\label{fig3}
\end{figure}

\begin{figure}[p]
\centering
\includegraphics[scale=0.4]{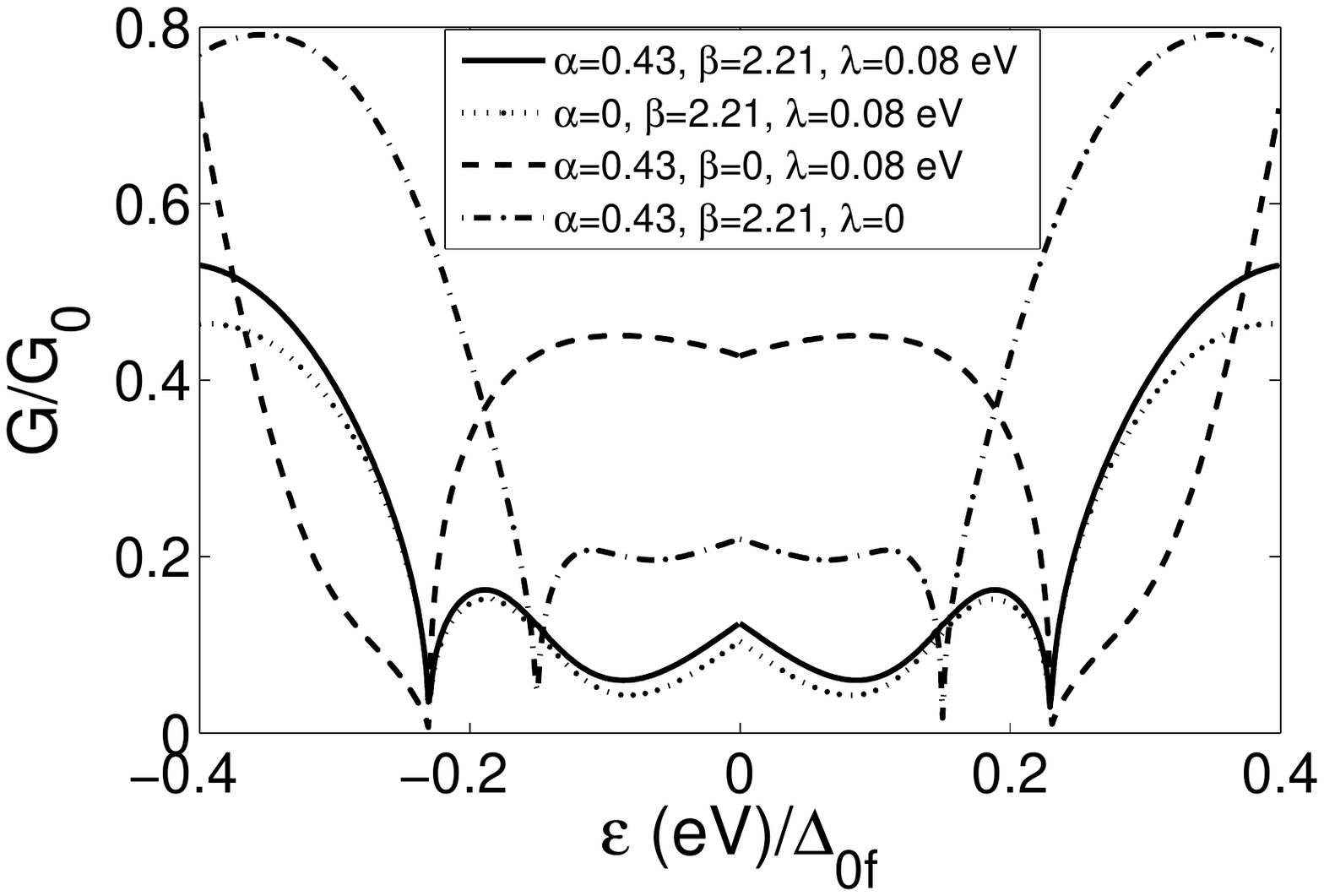}
\caption{}
\label{fig4}
\end{figure}

\begin{figure}[p]
\centering
\includegraphics[scale=0.4]{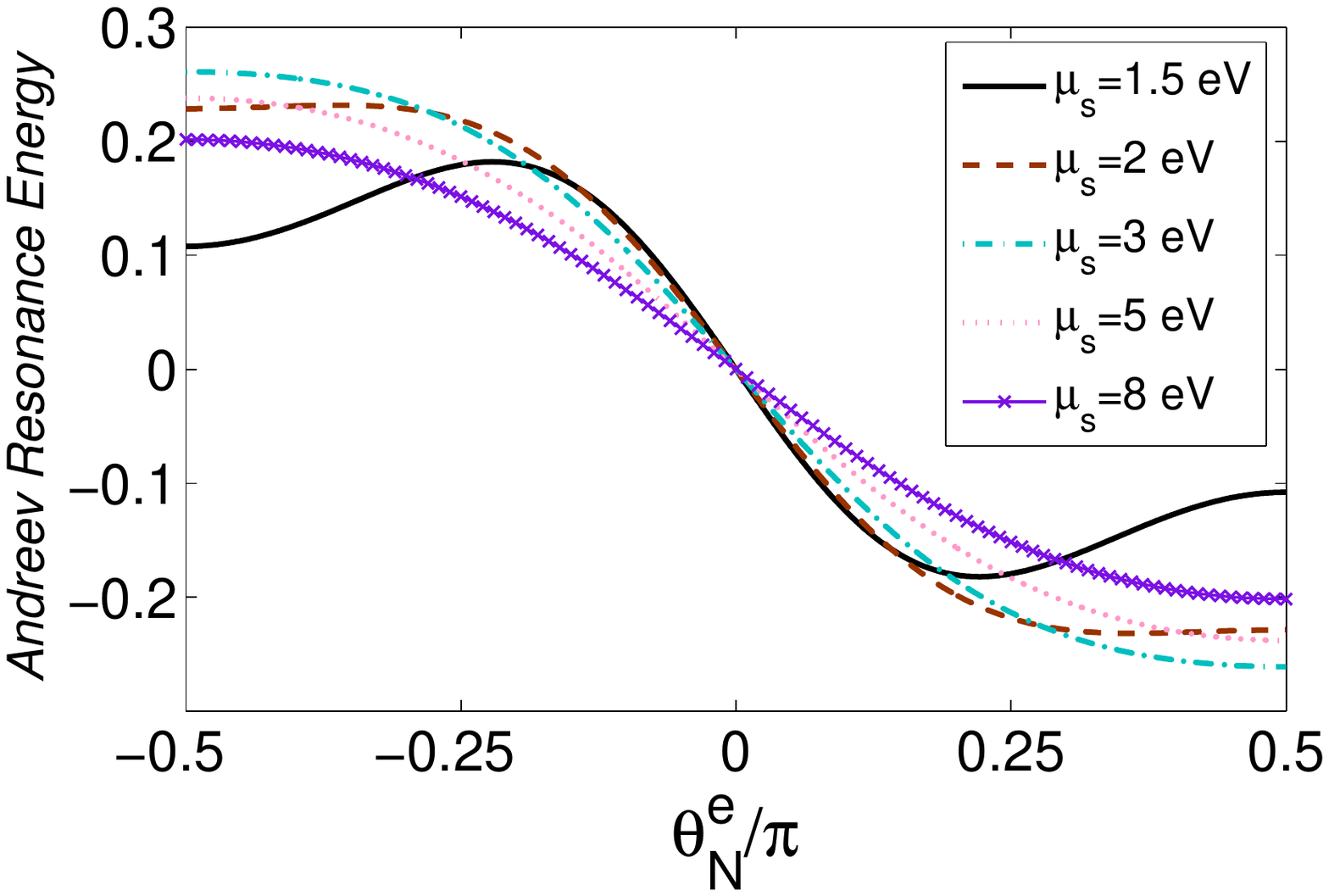}
\caption{}
\label{fig5}
\end{figure}

\begin{figure}[ht]
\centering
\includegraphics[scale=0.4]{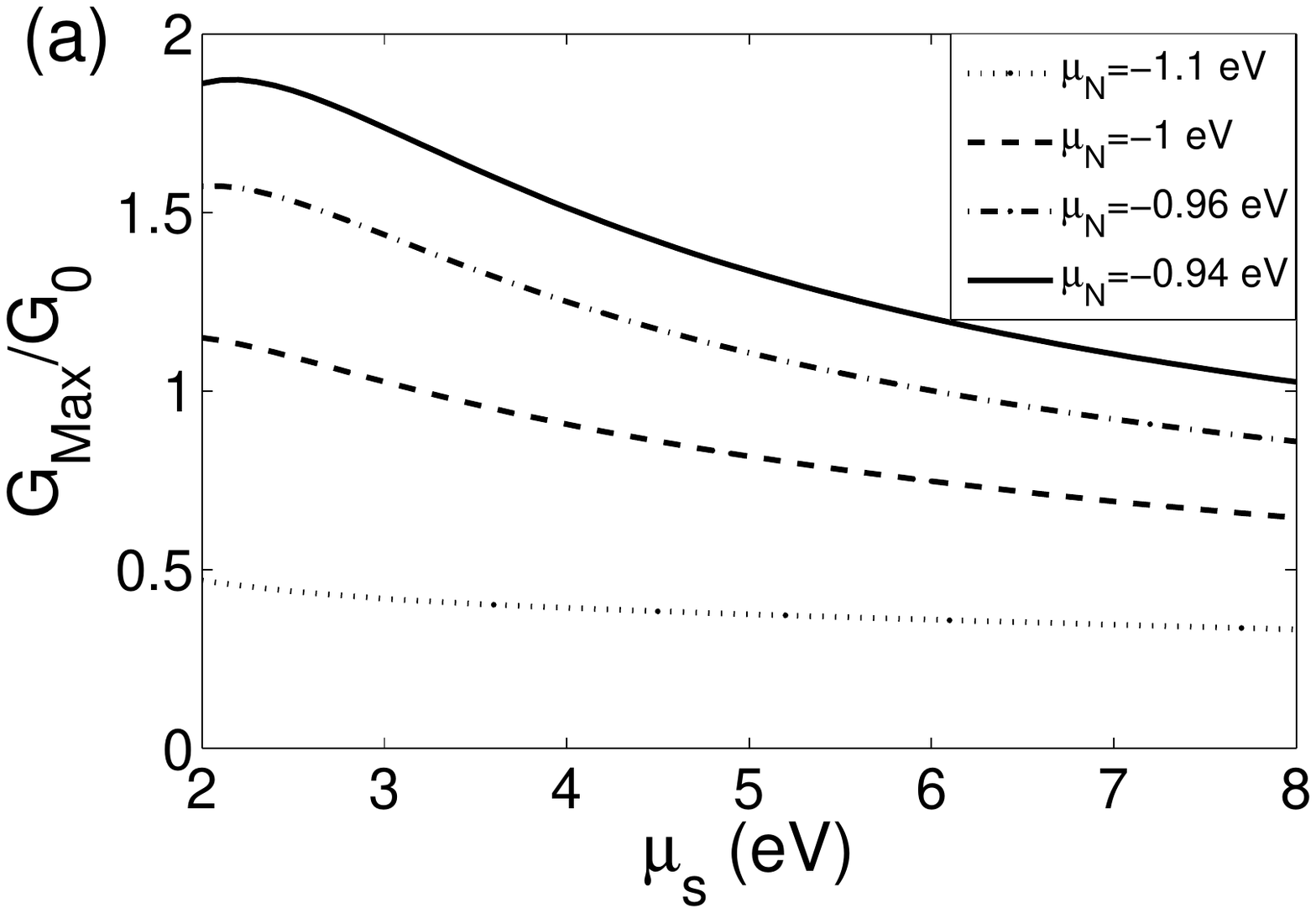}
\end{figure}

\begin{figure}[ht]
\centering
\includegraphics[scale=0.4]{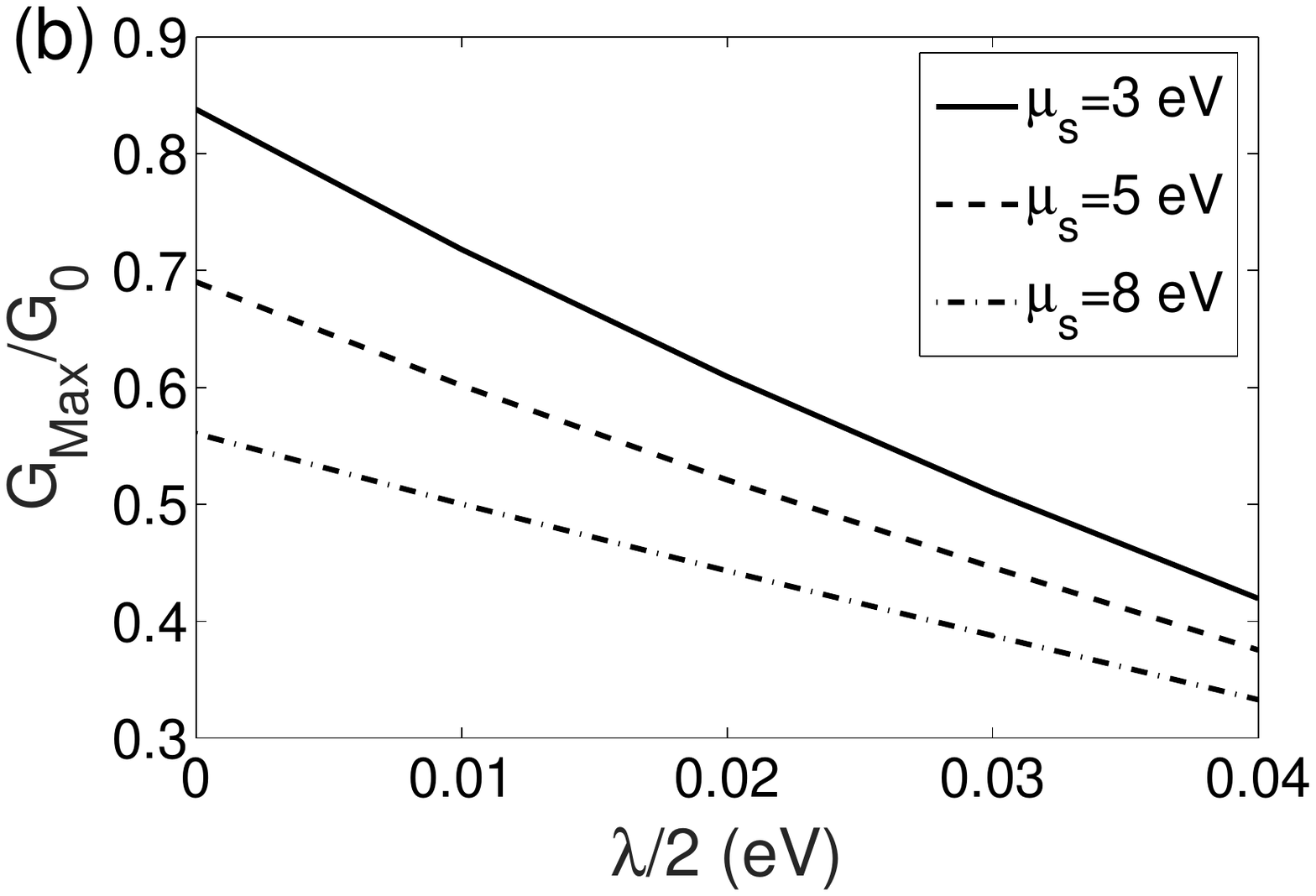}
\end{figure}

\begin{figure}[ht]
\centering
\includegraphics[scale=0.4]{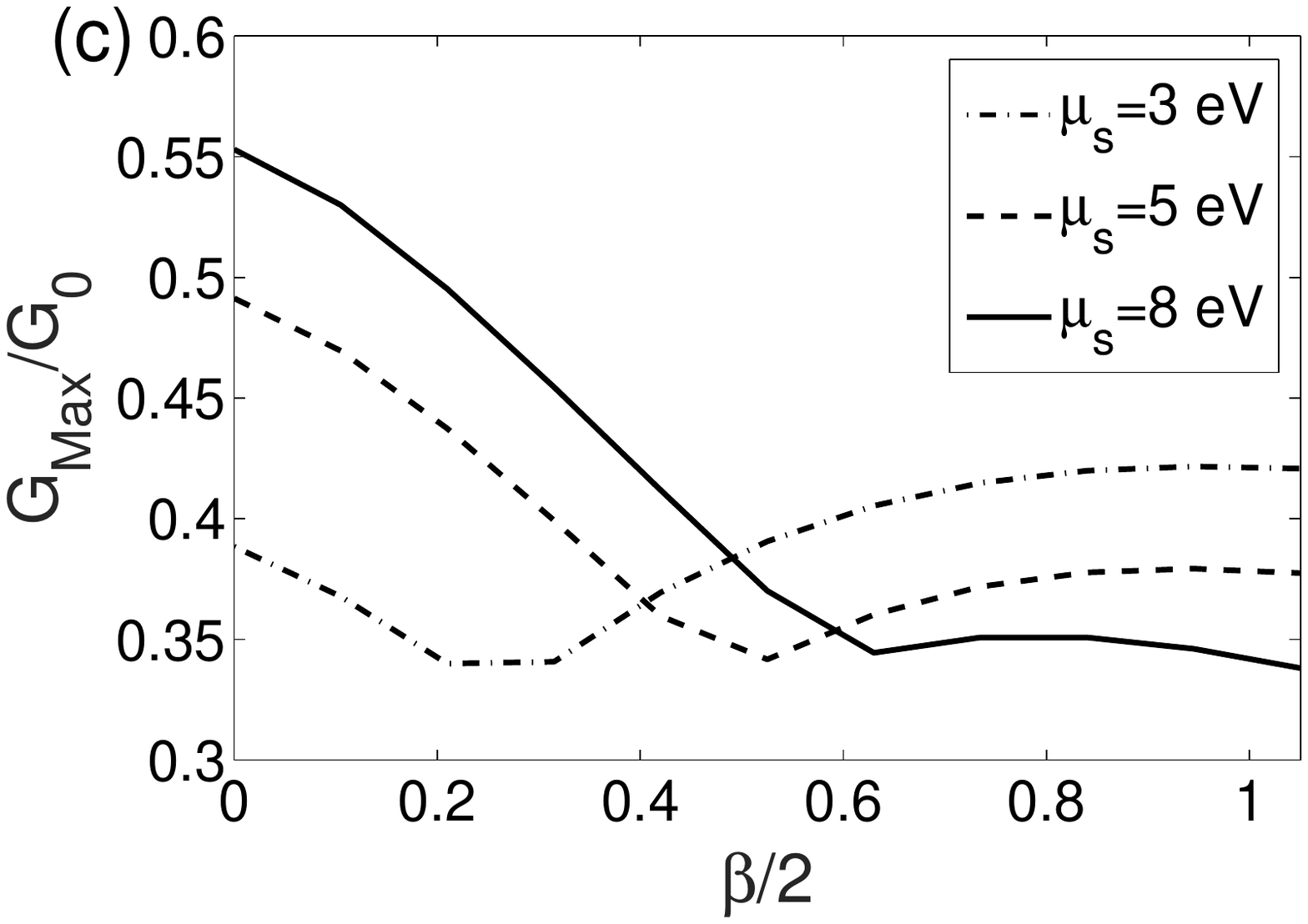}
\caption{(a),(b),(c)}
\label{fig6}
\end{figure}

\end{document}